\documentclass[twocolumn,notitlepage,prx,superscriptaddress,longbibliography]{revtex4-2}

\usepackage{listings}
\usepackage{tikz}
\usepackage{times}
\usepackage{graphicx}
\usepackage{float}
\usepackage{latexsym,amsmath,amssymb,bm,euscript}
\usepackage{color}
\usepackage{subfigure}
\usepackage{epstopdf}
\usepackage[colorlinks=true,linkcolor=blue,citecolor=blue]{hyperref}
\usepackage{soul}
\usepackage[normalem]{ulem}
\usepackage{mathrsfs}
\usepackage{amssymb}
\usepackage{algpseudocode}
\usepackage{algorithm}
\usepackage{amsmath}
\usepackage{braket}
\usepackage{booktabs}
\usepackage{chngcntr}
\usepackage{lettrine}
\usepackage{bbding}
\usepackage{xspace}
\usepackage{textcomp}
\usepackage{textcase}
\usepackage{setspace}
\usepackage[T1]{fontenc}
\usepackage{float}
\usepackage{graphicx}
\usepackage{multirow}
\usepackage{mathtools}
\usepackage{threeparttable}
\usepackage[vlined, ruled, algo2e, linesnumbered]{algorithm2e}
\usepackage{changes}
\usepackage{bibunits}
\usetikzlibrary{shapes}

\def\para{\ensuremath{/\kern -0.8em /}\xspace}
\def\ket#1{|{#1}\rangle}                 %% |#1>
\def\beqn{\begin{eqnarray}}
	\def\eeqn{\end{eqnarray}}
\def\beq{\begin{equation}}
	\def\eeq{\end{equation}}

\newcommand{\Beq}{\begin{eqnarray*} }
	\newcommand{\Eeq}{\end{eqnarray*} }
\newcommand{\Bmat}{\left(\begin{matrix}}
	\newcommand{\Emat}{\end{matrix}\right)}

\graphicspath{{../Fig/}}

\begin{document}
	\title{Giant Magnetocaloric Effect in Spin Supersolid Candidate Na$_2$BaCo(PO$_4$)$_2$}
	
	\author{Junsen Xiang}
	\thanks{These authors contributed equally to this work.}
	\affiliation{Beijing National Laboratory for Condensed Matter Physics, 
		Institute of Physics, Chinese Academy of Sciences, Beijing 100190, China}
	
	\author{Chuandi Zhang}
	\thanks{These authors contributed equally to this work.}
	\affiliation{School of Physics, Beihang University, Beijing 100191, China}
	
	\author{Yuan Gao}
	\thanks{These authors contributed equally to this work.}
	\affiliation{School of Physics, Beihang University, Beijing 100191, China}
	\affiliation{CAS Key Laboratory of Theoretical Physics, Institute of 
		Theoretical Physics, Chinese Academy of Sciences, Beijing 100190, China}
	
	\author{Wolfang Schmidt}
	\affiliation{Jülich Centre for Neutron Science JCNS at Institut Laue-Langevin (ILL), 
		Forschungszentrum Jülich GmbH, Boite Postale 156, 38042 Grenoble Cedex 9, France}
	
	\author{Karin Schmalzl}
	\affiliation{Jülich Centre for Neutron Science JCNS at Institut Laue-Langevin (ILL), 
		Forschungszentrum Jülich GmbH, Boite Postale 156, 38042 Grenoble Cedex 9, France}
	
	\author{Chin-Wei Wang}
	\affiliation{Australian Nuclear Science and Technology Organisation, Lucas Heights NSW 2234, Australia}
	
	\author{Bo Li}
	\affiliation{School of Physics, Beihang University, Beijing 100191, China}
	
	\author{Ning Xi}
	\affiliation{CAS Key Laboratory of Theoretical Physics, Institute of Theoretical 
		Physics, Chinese Academy of Sciences, Beijing 100190, China}
	
	\author{Xin-Yang Liu}
	\affiliation{School of Physics, Beihang University, Beijing 100191, China}
	\affiliation{CAS Key Laboratory of Theoretical Physics, Institute of 
		Theoretical Physics, Chinese Academy of Sciences, Beijing 100190, China}
	
	\author{Hai Jin}
	\affiliation{Department of Astronomy, Tsinghua University, Beijing 100084, China} 
	
	\author{Gang Li}
	\affiliation{Beijing National Laboratory for Condensed Matter Physics, 
		Institute of Physics, Chinese Academy of Sciences, Beijing 100190, China}
	
	\author{Jun Shen}
	\affiliation{Technical Institute of Physics and Chemistry, Chinese Academy 
		of Sciences, Beijing 100190, China}
	
	\author{Ziyu Chen}
	\affiliation{School of Physics, Beihang University, Beijing 100191, China}
	
	\author{{Yang Qi}}
	\affiliation{State Key Laboratory of Surface Physics and Department of Physics, 
		Fudan University, Shanghai 200433, China}
	
	\author{Yuan Wan}
	\affiliation{Beijing National Laboratory for Condensed Matter Physics, 
		Institute of Physics, Chinese Academy of Sciences, Beijing 100190, China}
	
	\author{Wentao Jin}
	\email{wtjin@buaa.edu.cn}
	\affiliation{School of Physics, Beihang University, Beijing 100191, China}
	
	\author{Wei Li}
	\email{w.li@itp.ac.cn}
	\affiliation{CAS Key Laboratory of Theoretical Physics, Institute of 
		Theoretical Physics, Chinese Academy of Sciences, Beijing 100190, China}
	\affiliation{CAS Center for Excellence in Topological Quantum Computation, 
		University of Chinese Academy of Sciences, Beijng 100190, China}
	\affiliation{Peng Huanwu Collaborative Center for Research and Education, 
		Beihang University, Beijing 100191, China}
	
	\author{Peijie Sun}
	\email{pjsun@iphy.ac.cn}
	\affiliation{Beijing National Laboratory for Condensed Matter Physics, 
		Institute of Physics, Chinese Academy of Sciences, Beijing 100190, China}
	
	\author{Gang Su}
	\email{gsu@ucas.ac.cn}
	\affiliation{Kavli Institute for Theoretical Sciences, and 
		School of Physical Sciences, University of Chinese 
		Academy of Sciences, Beijng 100049, China}
	\affiliation{CAS Center for Excellence in Topological 
		Quantum Computation, University of Chinese Academy 
		of Sciences, Beijng 100190, China}
	
	\begin{abstract}
		
		\noindent{\bf 
			Supersolid, an exotic quantum state of matter that consists of particles 
			forming an incompressible solid structure while simultaneously showing 
			superfluidity of zero viscosity~\cite{Leggett1970}, is one of the long-standing 
			pursuits in fundamental research~\cite{Kim2004,SSColloquium2012}. 
			Although the initial report of $^4$He supersolid turned out to be an 
			artifact~\cite{Kim2012}, this intriguing quantum matter has inspired 
			enthusiastic investigations into ultracold quantum gases~\cite{Li2017Nature,
				Leonard2017Nature,Tanzi2019Nature,Norcia2021Nature}. Nevertheless, 
			the realization of supersolidity in condensed matter remains elusive. 
			Here we find evidence for a quantum magnetic analogue of supersolid --- 
			the spin supersolid --- in the recently synthesized triangular-lattice 
			antiferromagnet Na$_2$BaCo(PO$_4$)$_2$~\cite{Zhong2019}. Notably, 
			a giant magnetocaloric effect related to the spin supersolidity is observed in 
			the demagnetization cooling process, manifesting itself as two prominent 
			valley-like regimes, with the lowest temperature attaining below 100 mK. 
			Not only is there an experimentally determined series of critical fields but 
			the demagnetization cooling profile also shows excellent agreement with the 
			theoretical simulations with an easy-axis Heisenberg model. Neutron 
			diffractions also successfully locate the proposed spin supersolid phases 
			by revealing the coexistence of three-sublattice spin solid order and interlayer 
			incommensurability indicative of the spin superfluidity. Thus, our results indicate 
			a strong entropic effect of the spin supersolid phase in a frustrated quantum 
			magnet and open up a viable and promising avenue for applications in sub-Kelvin 
			refrigeration, especially in the context of persistent concerns about helium 
			shortages~\cite{Cho2009Science,Kramer2019Helium}.
		}
	\end{abstract}
	
	\date{\today}
	\maketitle
	\noindent
	The pursuit of exotic states of matter with novel quantum orders 
	constitutes a major theme in modern condensed matter physics. 
	As a solid-state matter that simultaneously exhibits superfluid 
	order, supersolid has no classical counterpart~\cite{Kim2004,
		SSColloquium2012,Kim2012,Li2017Nature,Leonard2017Nature,
		Tanzi2019Nature,Norcia2021Nature}. Despite the difficulty in 
	searching for supersolidity in $^4$He~\cite{Kim2004,Kim2012}, 
	there has been a clear theoretical prediction for the emergence of 
	supersolidity on a triangular lattice with hardcore bosons~\cite{Melko2005,
		Wessel2005,Heidarian2005,Prokofev2005} or equivalently $S=1/2$ 
	quantum spins~\cite{Heidarian2010,Yamamoto2014,Yamamoto2015,
		Sellmann2015}. The latter constitutes a quantum magnetic analogue 
	of supersolid, where both the lattice translational and spin rotational 
	symmetries are broken simultaneously. 
	Spin supersolid has also been conjectured for a three-dimensional 
	($3$D) high-spin chromium spinel compound MnCr$_2$S$_4$ (MCS) 
	based on symmetry arguments and an analogy to the quantum lattice 
	gas model~\cite{Matsuda1970,Liu1973,Loidl2017}. However, the precise 
	quantum spin model of MCS and a comprehensive understanding of 
	its phase diagram are still elusive. Therefore, a conclusive prototype 
	system of supersolid remains to be discovered in experiments. 
	
	The highly frustrated triangular-lattice antiferromagnets (TLAFs) are a fertile 
	ground for breeding novel quantum spin states~\cite{Anderson1973, 
		Collins1997,Starykh2015,Yutaka2012PRL,Zhou2012}. Among them, 
	the spin-$1/2$ Co-based equilateral TLAF Na$_2$BaCo(PO$_4$)$_2$ 
	(NBCP)~\cite{Zhong2019}, with unique easy-axis anisotropy, has garnered 
	significant interest~\cite{LiN2020,Lee2021,Wellm2021,Gao2022QMats} 
	along two conflicting lines: quantum spin liquid~\cite{Zhong2019,Lee2021} 
	and magnetically ordered state~\cite{LiN2020}.
	Recently, a nearly ideal easy-axis TLAF model description of NBCP has been 
	put forward, which reconciles the divergent experimental observations within 
	a coherent picture~\cite{Gao2022QMats}. Remarkably, the system was 
	predicted to host spin supersolid states under both zero and finite magnetic 
	fields, which raises intriguing proposals for finding this exotic quantum state.
	
	% ====== Fig. 1: Illustration  ====== %
	\begin{figure*}[htp]
		\includegraphics[width=1\linewidth]{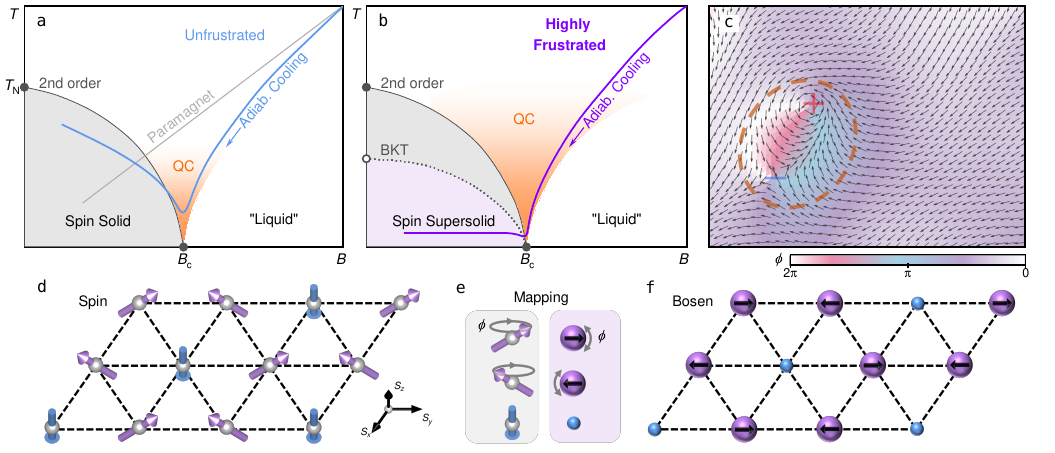}
		\caption{\textbf{Illustration of spin supersolid cooling, 
				spin-boson mapping, and the U(1) phase fluctuations.} 
			\textbf{a} Cooling effect near the conventional solid-liquid quantum 
			phase transition, compared to \textbf{b} that near the supersolid-liquid 
			transition in a triangular-lattice magnet. The adiabatic demagnetization
			curve of a paramagnetic coolant follows the ideal linear trace with a 
			constant $T/B$ ratio, while both quantum magnets exhibit enhanced 
			cooling effects in the quantum critical (QC) region near the QCP. However, 
			the isentropic line of the ordered magnet in \textbf{a} generally bounces 
			back for $B < B_{\rm c}$ after crossing the QCP, while the spin supersolid 
			in \textbf{b} shows a very distinct flat valley structure.
			\textbf{{c}} The arrows (and colors) denote phase $\phi$ on one 
			of the three sublattices (to be explained below), obtained from the 
			Monte Carlo calculations on a triangular lattice XY model.
			\textbf{{d-f}} show the spin-boson mapping where the 
			Hilbert space of  $S=1/2$ spins can be mapped exactly into hardcore 
			bosons: $\ket{\uparrow} \rightarrow \ket{1}, \, \ket{\downarrow} 
			\rightarrow \ket{0}$, i.e., the spin-up (-down) state corresponds to the 
			occupied (unoccupied) boson site {(shown in \textbf{e})}. 
			\textbf{{d}} The supersolid state (Y-state) can be mapped into \textbf{{f}} 
			hardcore bosons with a charge density wave pattern and an additional 
			in-plane phase $\phi$ [U($1$) phase] denoted by the bold arrows. 
			The U($1$) phase fluctuations in \textbf{{d}}
			or \textbf{{f}} are illustrated in \textbf{{c}}, where the long wave-length 
			harmonic fluctuations (Goldstone mode) and topological excitations 
			[vortex (``+'')-antivortex (``-'') pair] are indicated.} 
		\label{Fig1}
	\end{figure*}
	
	Nevertheless, probing the spin supersolid at very low temperatures 
	faces significant experimental challenges. One of the key properties 
	underlying the novel phase is the entropy landscape. For quantum 
	magnets, it can be obtained by measuring the magnetocaloric effect 
	(MCE), which represents an adiabatic temperature change as field varying. 
	Conventionally, the MCE is most pronounced near the finite-$T$ magnetic 
	phase transitions~\cite{Weiss1917}. At temperatures approaching absolute 
	zero, magnetocaloric effect (MCE) can also be greatly enhanced in regions 
	near strongly fluctuating states and quantum critical points (QCPs)
	\cite{Wolf2011}. The magnetic Gr\"uneisen parameter $\Gamma_B$ 
	that characterizes the MCE properties exhibits an abrupt sign change 
	and diverges across the QCP~\cite{Zhu2003,Tokiwa2009,Gegenwart2016}. 
	MCE measurements can thus be used to sensitively detect QCPs and strongly 
	fluctuating spin states, allowing for the mapping out of the phase diagram
	\cite{Rost2009,Gegenwart2016,Wang2018MCE}.
	
	The significant MCE related to fluctuating spin states~\cite{Zhu2003,Wolf2011,
		Wolf2016} also makes quantum magnets promising coolants for helium-free 
	sub-Kelvin refrigeration, which becomes increasingly important for space 
	applications~\cite{Shirron2014} and quantum technologies~\cite{Jahromi2019nasa}. 
	The conventional paramagnetic cooling~\cite{Giauque1933} follows a straight 
	$T$-$B$ line with a constant slope. Distinctly, in a quantum magnet, 
	the temperature drops quickly near a QCP (c.f., Fig.~\ref{Fig1}\textbf{a}), 
	which, however, rises up again in the lower-field side ($B<B_{\rm c}$)
	\cite{Rost2009,Wang2018MCE}. Therefore, searching for quantum spin 
	systems with more pronounced and persistent cooling effect (c.f., Fig.~\ref{Fig1}\textbf{b}) 
	is of both fundamental and practical significance.
	
	% ====== Fig. 2: Magnetic Cooling ====== %
	\begin{figure*}[htp]
		\includegraphics[width=0.725\linewidth]{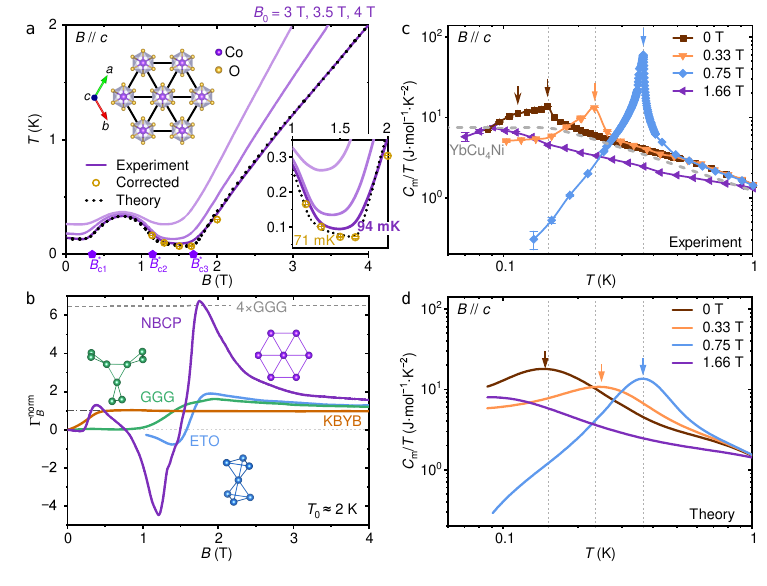}
		\caption{\textbf{Quasi-adiabatic demagnetization cooling and low-temperature 
				heat capacity of Na$_2$BaCo(PO$_4$)$_2$.} \textbf{a} Measured adiabatic
			cooling from an initial temperature $T_0=2$~K and various peak fields 
			$B_0=3$~T, $3.5$~T, and $4$~T. The corrected results based on the 
			experimental entropy data (Extended Data Fig.~\ref{EDFig4}) and simulated 
			results of the easy-axis TLAF model virtually coincide, and both can reach
			approximately $71$~mK near $B_{\rm c3}^*$ (see the inset at the bottom 
			right). The error bars of the corrected results are estimated from the fitting 
			uncertainties in nuclear contributions. The triangular lattice layer of CoO$_6$ 
			in NBCP is sketched in the (upper left) inset, with the crystalline $a, b$, 
			and $c$ axes also indicated. The purple pentagons on the horizontal axis 
			represent the determined field-induced QCPs ($B_{\rm c1}^*$, $B_{\rm c2}^*$, 
			$B_{\rm c3}^*$). \textbf{b} Normalized Gr\"uneisen parameters 
			$\Gamma_B^{\rm norm}$ of NBCP (with $B_0=4$~T) compared to other 
			frustrated magnets from the same initial temperature $T_0 \simeq 2$~K, e.g., 
			GGG ($B_0=4$~T), ETO ($B_0=4$~T)~\cite{Wolf2016}, 
			and KBYB ($B_0=5$~T)~\cite{Tokiwa2021}. 
			The horizontal dash-dotted line of $\Gamma_B^{\rm norm} \equiv 1$
			represents the behavior of an ideal paramagnet, and the geometrically
			frustrated structures in GGG, ETO, and NBCP are also depicted as insets. 
			\textbf{c}, \textbf{d} Measured and simulated specific heat curves of NBCP 
			under field \textit{B} // \textit{c}, with the arrows indicating the $C_{\rm m}/T$ 
			peaks in the experimental results and the model calculations. Despite 
			the less pronounced $C_{\rm m}/T$ peaks due to the limited system 
			sizes, the theoretical calculations agree with the experimental results. 
			The grey dashed curve in \textbf{c} plots $C_{\rm m}/T$ of heavy-fermion 
			metal YbCu$_4$Ni as a comparison~\cite{shimura2022magnetic}.
		} 
		\label{Fig2}
	\end{figure*}
	% ================= %
	
	\bigskip 
	\noindent 
	\textit{\bf Results: Triangular lattice compound} \\
	We synthesize high quality single-crystal samples of NBCP for MCE 
	and neutron diffraction studies, where the Co$^{2+}$ ions constitute 
	an effective $S=1/2$ spins (Methods) on a perfect triangular lattice (see 
	$ab$ plane in inset of Fig.~\ref{Fig2}\textbf{a}). The magnetic interactions 
	between the spin-orbit moments of Co$^{2+}$ ions follow the $D_{\rm 3d}$ 
	site symmetry. Specific heat measurements of NBCP reveal a peak at 
	approximately $150$~mK~\cite{LiN2020,Huang2022NBCP}, while the 
	muon spin relaxation finds no signature of static magnetism down to 
	$80$~mK~\cite{Lee2021}. Despite a relatively high density of magnetic 
	ions of $5.8\times10^{21}$ cm$^{-3}$, the nearest-neighbor magnetic 
	interactions are of moderate strength ($1$-$2$~K) {between} the sites 
	via the super-super exchange~\cite{Zhong2019,LiN2020,Lee2021,
		Wellm2021,Gao2022QMats}. 
	
	Through many-body simulations (Methods) of the easy-axis TLAF model for 
	NBCP~\cite{Gao2022QMats}, it has been predicted that under an out-of-plane
	field ($B$//$c$), an up-up-down (UUD) spin solid and two spin supersolid phases
	(Y- and V-states) can emerge. The latter phases spontaneously break the 
	lattice translational and U($1$) spin rotational symmetries~\cite{Yamamoto2014,
		Yamamoto2015,Sellmann2015,Gao2022QMats}.
	Through the exact hardcore boson mapping (Figs.~\ref{Fig1}\textbf{d-f} 
	and Methods), the UUD state corresponds to a gapped boson Mott insulator 
	with fractional ($2/3$) filling, evidenced by the $1/3$-magnetization plateau 
	integrated from the low-temperature ac susceptibility measurements~\cite{LiN2020}. 
	On the other hand, the Y- and V-states are supersolid states that exhibit both 
	charge density wave and superfluid orders, and remain to be 
	explored experimentally.
	\\

	% ======== Fig. 3: Low-T MCE measurements and QCPs ======= %
	\begin{figure*}[thp]
		\includegraphics[width=0.75\linewidth]{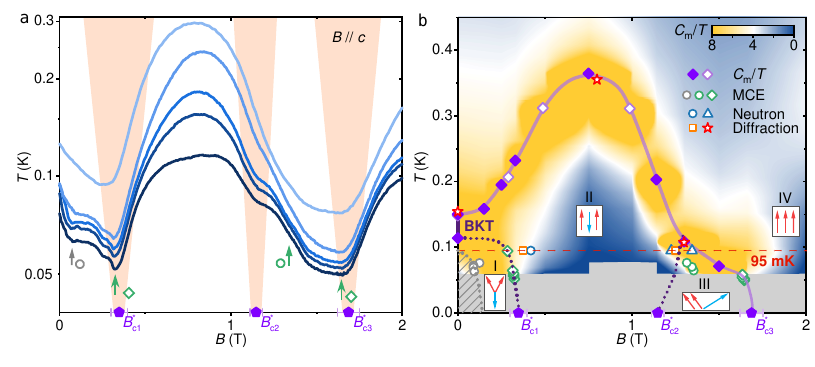}
		\caption{\textbf{Low-temperature MCE, field-induced quantum 
				phase transitions, and the field-temperature phase diagram.} 
			\textbf{a} Quasi-adiabatic cooling starting from various initial temperature 
			$T_0 \leq 300$~mK and a fixed initial field ($B_0=3$~T). There are quantum 
			critical fans that emanate from the QCPs at $B_{\rm c1}^*, B_{\rm c2}^*$, 
			and $B_{\rm c3}^*$ (see Extended Data Fig.~\ref{EDFig7}). The abrupt 
			changes in temperature indicated by the gray arrows (forming a grey 
			hatched regime in \textbf{b}) are possibly due to $3$D couplings. 
			\textbf{b} There are four primary phases (I to IV; see the main text) in the 
			phase diagram. The topmost phase boundary denotes the melting of the 
			spin solid order, as determined from the $C_{\rm m}$ data (filled rhombi in 
			the present work and open purple rhombi from Ref.~\cite{LiN2020}) and neutron 
			diffraction (open asterisks, determined from  Fig.~\ref{Fig4}\textbf{e}). The green 
			open circles denote the location indicated by the green arrow in \textbf{a} where 
			rapid temperature change occurs between $B_{\rm c2}^*$ and $B_{\rm c3}^*$ 
			in the $T$-$B$ curves, and the green open rhombi are determined from the 
			minimal cooling temperature near $B_{\rm c1}^*$ and $B_{\rm c3}^*$. 
			The purple dotted lines indicate the BKT transitions, and the thick purple 
			solid line along the vertical axis represents the BKT phase with an algebraic 
			solid spin order under zero field. The critical fields obtained from the field 
			dependence of neutron diffractions are conducted at $95$~mK and shown
			with the same symbol code as Fig.~\ref{Fig4}\textbf{f} [e.g., orange open 
			square for ($1/3,1/3,1$) peak].
		}
		\label{Fig3} 
	\end{figure*}
	% ================= %
	
	\noindent
	{\bf Spin supersolid cooling} \\
	As shown in Fig.~\ref{Fig2}\textbf{a}, the NBCP sample cools down 
	to as low as $94$~mK (near $1.5$~T) through a quasi-adiabatic 
	demagnetization process from the initial conditions of $T_0=2$~K 
	and $B_0=4$~T. Besides, there are two valley-like regimes, i.e., $B\leq 
	B_{\rm c1}^*\simeq 0.35$~T and $B_{\rm c2}^* \leq B \leq B_{\rm c3}^*$ 
	(with $B_{\rm c2}^* \simeq 1.15$~T and $B_{\rm c3}^*\simeq1.69$~T), 
	where the sample temperature greatly decreases and remains 
	persistently at low values.
	
	The remarkable observation in Fig.~\ref{Fig2}\textbf{a} can be 
	quantitatively explained with the easy-axis TLAF model, where 
	an excellent agreement between the experimental results and 
	the computed adiabatic cooling curves can be observed. Scrutinizing 
	the inset of Fig.~\ref{Fig2}\textbf{a}, we find the attained lowest 
	temperature ($94$~mK) is still slightly higher than the simulated 
	value of about $71$~mK. The discrepancy between the measured 
	and theoretical transition temperatures is likely due to the inevitable 
	heat leakage in our quasi-adiabatic cooling stage (Methods and 
	Extended Data Fig.~\ref{EDFig3}). To account for this, we have 
	corrected the adiabatic cooling temperature using the entropy data 
	integrated from the measured specific heat, rendering a remarkable 
	agreement in Fig.~\ref{Fig2}\textbf{a}. 
	
	Given the aforementioned agreement, it is natural to propose the 
	concept of \textit{spin supersolid cooling}. As illustrated in Fig.~\ref{Fig1}\textbf{b}, 
	when the applied magnetic fields drive the system from correlated 
	paramagnetic (liquid) spin states into the supersolid phase, giant 
	and persistent magnetocaloric responses occur. As NBCP is ideally 
	of $2$D nature~\cite{LiN2020,Wellm2021}, the superfluid order is 
	quasi long-range at finite temperature according to the 
	Berezinskii–Kosterlitz–Thouless (BKT) theory~\cite{Berezinskii1971,
		Kosterlitz1973}. There are low-energy excitations, e.g., Goldstone 
	mode related to the U(1) phase fluctuations (c.f., Figs.~\ref{Fig1}\textbf{c-f}), 
	accounting for the tremendous magnetic entropy even at very low temperature.
	{In contrast, the temperature rises up and forms a ridge-like shape 
		in the $T$-$B$ curve for intermediate fields, i.e., $B_{\rm c1}^* 
		\leq B \leq B_{\rm c2}^*$~\cite{LiN2020,Gao2022QMats}, reflecting 
		the absence of low-energy fluctuations in the UUD spin solid phase.}
	
	To further quantify the MCE, we adopt the normalized magnetic 
	Gr\"uneisen parameter $\Gamma^{\rm norm}_B = \Gamma_B/
	\Gamma_B^0 =B \, \Gamma_B$, with $\Gamma_B=\frac{1}{T} 
	(\frac{\partial T}{\partial B})_S$ for the frustrated magnet and 
	$\Gamma_B^0 = \frac{1}{B}$ for the ideal paramagnet. In Fig.
	\ref{Fig2}\textbf{b}, we compare $\Gamma^{\rm norm}_B$ 
	between NBCP and several frustrated magnets including 
	Gd$_3$Ga$_5$O$_{12}$ (GGG), Er$_2$Ti$_2$O$_7$ (ETO)
	\cite{Wolf2016}, and KBaYb(BO$_3$)$_2$ (KBYB)~\cite{Tokiwa2021}. 
	We find pronounced peak-dip structures in the $\Gamma^{\rm norm}_B$ 
	curves near QCP in GGG, ETO, and NBCP, while KBYB remains unity 
	down to a very low field until it decreases towards zero. In particular, 
	the peak height of NBCP is more than four times the value for GGG. 
	This giant peak in $\Gamma_B^{\rm norm}$ clearly surpasses 
	that of other frustrated magnets, which we ascribe to the strong 
	fluctuations of both spin solid and superfluid orders near the supersolid 
	QCP at $B_{\rm c3}$. The spin-wave excitations fall into gapless 
	quadratic dispersions near $B_{\rm c3}$ with large density of states 
	(Extended Data Fig.~\ref{EDFig5}), which also naturally explains 
	the huge entropic effect.
	
	In practical quasi-adiabatic cooling measurements starting from the 
	same initial conditions in PPMS-based environments, we find that 
	NBCP can indeed reach a significantly lower temperature ($94$
	mK) than GGG ($365$~mK), ETO ($360$~mK)~\cite{Wolf2016}, 
	and the widely used paramagnetic coolants Fe(SO$_4$)$_2$(NH$_4$)$_2$ 
	$\cdot 6$H$_2$O ($124$~mK) and CrK(SO$_4$)$_2\cdot12$H$_2$O 
	($118$~mK) (Methods and Extended Data Fig.~\ref{EDFig6}). 
	Moreover, more detailed comparisons show that NBCP can absorb 
	more heat per unit volume and stay cooler for longer time. These results 
	qualify NBCP as a superior sub-Kelvin refrigerant with excellent cooling 
	performances.
	\\
	
	\begin{figure*}[htp]
		\includegraphics[width=1\linewidth]{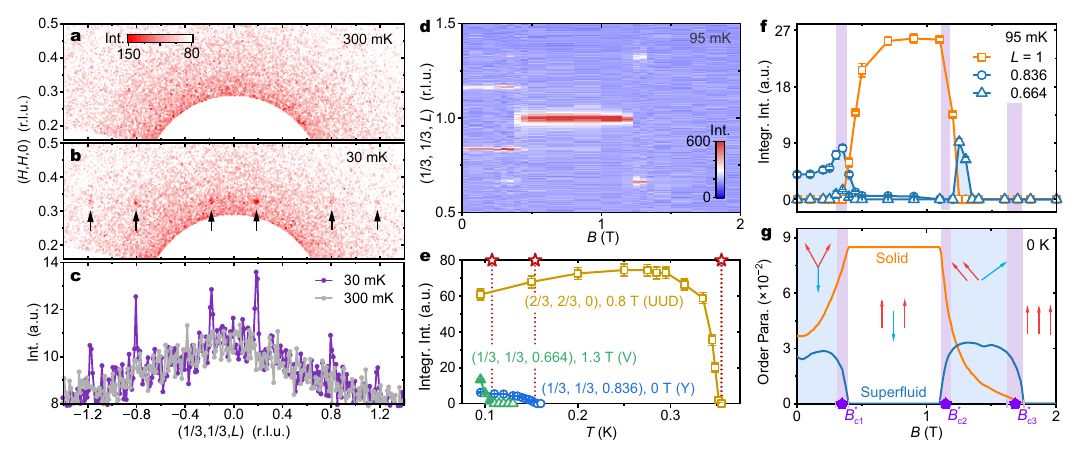}
		\centering
		\caption{{\textbf{Low-temperature neutron diffraction.} Zero-field neutron 
				diffraction patterns collected at $300$~mK and $30$~mK are shown in 
				\textbf{a} and \textbf{b}, respectively, where bright spots in \textbf{b} 
				represent spin ordering. \textbf{c} represents the diffraction intensities 
				cut along ($1/3$, $1/3$, $L$) direction, in which $H$ is integrated within 
				the range of [$0.31$, $0.34$]. 
				\textbf{d} shows reciprocal-space scans at $95$~mK under different fields 
				applied along the $c$ axis. The temperature dependencies of the integrated 
				intensities of ($1/3$, $1/3$, $0.836$), ($2/3$, $2/3$, $0$), and ($1/3$, $1/3$, $0.664$) 
				reflections measured in different fields are shown in \textbf{e}, where the 
				vertical dashed lines and red open stars denote the onset temperature of 
				the Y, UUD, and V phases, respectively. The integrated intensities of 
				($1/3$, $1/3$, $0.836$), ($1/3$, $1/3$, $1$), and ($1/3$, $1/3$, $0.664$) reflections 
				are plotted in \textbf{f} as a function of applied field. Error bars represent 
				standard deviations in \textbf{e} and \textbf{f}. In \textbf{g}, the ground-state  
				calculations of the TLAF model reveal the Y ($\nwarrow \nearrow 
				\downarrow$), UUD ($\uparrow \uparrow \downarrow$), V ($\nwarrow 
				\nwarrow \nearrow$), and the polarized ($\uparrow \uparrow \uparrow$) 
				phases,  with the simulated spin solid and superfluid order parameters 
				also shown in each phase. r.l.u.,reciprocal lattice units; a.u., arbitrary units.
		}}
		\label{Fig4}
	\end{figure*}

	\noindent
	{\bf Specific heat results} \\
	In Fig.~\ref{Fig2}\textbf{c}, we show the magnetic specific heat ($C_{\rm m}/T$) 
	under various fields and down to about $60$ mK (Methods) to further comprehend 
	the valley-like behavior in the {$T$-$B$} curves. We find very large low-$T$ 
	specific heat in the two supersolid valleys, i.e., $B\leq B_{\rm c1}^*$ and 
	$B_{\rm c2}^* \leq B \leq B_{\rm c3}^*$, which reflects the strong low-energy 
	fluctuations. The measured values of $C_{\rm m}/T$ even exceed that for a 
	typical heavy-fermion metal YbCu$_4$Ni that also has outstanding low-$T$ 
	MCE properties~\cite{shimura2022magnetic}. The magnetothermal properties 
	are highly field-tunable in NBCP. By varying the field to $0.75$~T (i.e., in the 
	UUD phase), the low-$T$ specific heat becomes smaller than that in the 
	supersolid valleys by an order of magnitude.
	
	In Fig.~\ref{Fig2}\textbf{d}, we further plot the TLAF simulation results of 
	$C_{\rm m}/T$, where the large low-temperature $C_{\rm m}/T$ values of 
	the spin supersolid phase are in great agreement with experiments. For 
	example, in the cases of $B=0.33$~T and $0.75$~T, the $C_{\rm m}/T$ 
	peaks correspond to the $3$-state Potts transition. The former ($0.33$~T) 
	locates within the supersolid phase and has large low-$T$ specific heat 
	even below the solid-order Potts transition, while $C_{\rm m}/T$ of the 
	latter case ($0.75$~T) rapidly drops after entering the UUD spin solid 
	phase. The experiment and calculation results indicate that the quantum 
	spin states and low-energy excitations undergo dramatic changes crossing 
	the spin supersolid transition, while remaining field-insensitive within the 
	supersolid phase. It thus leads to the distinct magnetocaloric responses 
	illustrated in Fig.~\ref{Fig1}\textbf{b} and practically observed in 
	Fig.~\ref{Fig2}\textbf{a}.
	\\
	
	\noindent
	{\bf Field-temperature phase diagram} \\ 
	Although the spin supersolid phase in NBCP can be revealed only 
	at very low temperature, its observation relies on no extra cooling 
	resources, but the giant MCE of itself during the demagnetization 
	process --- dubbed as \textit{magnetocaloric bootstrapping}. In 
	Fig.~\ref{Fig3}\textbf{a}, even from deeply in the sub-Kelvin regime
	($T_0\simeq 90$~mK), the system can reach about $50$~mK, 
	providing a clear manifestation of the spin supersolid and 
	related QCP.
	
	Specifically, we observe clear minima (at $B^*_{\rm c1} \simeq 0.35$($5$)~T 
	and $B_{\rm c3}^*\simeq 1.69$($6$)~T) and shoulder-like features 
	(at $B_{\rm c2}^*\simeq 1.15$($4$)~T) in the quasi-adiabatic $T$-$B$ curves. 
	The locations of these QCPs can be corroborated by the Gr\"uneisen 
	parameter $\Gamma_B$ deduced from the $T$-$B$ curves (Extended 
	Data Fig.~\ref{EDFig7}). The three transition fields determined are in very 
	good agreement with previous susceptibility (down to $22$~mK) and 
	thermal conductivity (down to $92$~mK) measurements~\cite{LiN2020}. 
	
	As shown in Fig.~\ref{Fig3}\textbf{b}, collecting the critical fields determined 
	from multiple measurements, we map the NBCP phase diagram under fields 
	$B$ // $c$ axis. There exists a dome-like upper boundary separating the UUD 
	(II) and paramagnetic phase (IV). The phases I (Y-state) and III (V-state) 
	constitute the quantum spin analogue of supersolid states.
	The purple dotted lines in Fig.~\ref{Fig3}\textbf{b} represent the tentative BKT 
	phase boundaries determined from the MCE data and neutron diffractions 
	(see below), terminating at QCPs $B_{\rm c1}^*$ and $B_{\rm c2}^*$.
	\\
	
	\noindent
	{\bf Neutron diffraction} \\ 
	To provide microscopic evidence for the spin supersolid, neutron diffraction 
	measurements are conducted on a large single crystal of NBCP under zero 
	and finite fields (Methods). As shown in Figs.~\ref{Fig4}\textbf{a}
	and \textbf{b}, additional reflections can be observed in zero field upon cooling 
	from $300$ mK to $30$ mK, implying the occurrence of spin ordering in NBCP. 
	The diffraction patterns at the two temperatures are better visualized in the cut 
	along ($1/3$, $1/3$, $L$) direction as shown in Fig.~\ref{Fig4}\textbf{c}. A 
	magnetic propagation vector of ($1/3$, $1/3$, $0.183$($11$)) is revealed, where 
	the in-plane components clearly indicate the formation of a three-sublattice 
	ordering as expected for the spin supersolid state. 
	
	Reciprocal-space scans at $95$~mK under applied fields  along the $c$ axis 
	are presented in Fig.~\ref{Fig4}\textbf{d}, which reveals a significant change in 
	the ordering vector at the transition fields. In the regime $B \leq B^*_{\rm c1}$ 
	and $B^*_{\rm c2} \leq B < B^*_{\rm c3}$, the ordering vector locates at ($1/3$, 
	$1/3$, $q_{c}$) with an incommensurate out-of-plane $q_{c}$, while the system 
	exhibits commensurate ordering for $B^*_{\rm c1} \leq B \leq B^*_{\rm c2}$. 
	The observed incommensurate $q_{c}$ can be attributed to the sensitivity of 
	spin supersolid states to weak interlayer couplings, which can give rise to a $3$D 
	magnon condensation at $(1/3$, $1/3$, $q_{c}$). As evidenced in 
		Extended Data Fig.~\ref{EDFig8}, the almost resolution-limited peak widths of 
		the magnetic reflection suggests both long-range intra- and interlayer correlations. 
		This can be ascribed to the out-of-plane magnetic moments given their long-range 
		ordered nature, while the in-plane components are still strongly fluctuating within 
		the $2$D plane (Methods). In contrast, interlayer couplings have little influence on 
	the gapped UUD solid phase, making it distinct from the spin supersolid phase.
	
	In Fig.~\ref{Fig4}\textbf{f}, we show the diffraction intensities at ($1/3$, $1/3$, $0.836$), 
	($1/3$, $1/3$, $1$), and ($1/3$, $1/3$, $0.664$) under various fields, which show great 
	similarity with the density matrix renormalization group (DMRG, Methods) results 
	shown in Fig.~\ref{Fig4}\textbf{g}. In the supersolid phase (Y- and V-states), 
	intertwined spin solid and superfluid orders coexist, while in the UUD phase, 
	only a solid order is present. By comparing experiments with DMRG calculations, 
	we clearly identify that ($1/3$, $1/3$, $0.836$) and ($1/3$, $1/3$, $0.664$) correspond to 
	the Y- and V-states, respectively, while the ($1/3$, $1/3$, $1$) reflection corresponds 
	to the UUD order. 
	
	Besides, $B^*_{\rm c1}$ and $B^*_{\rm c2}$ determined from neutron measurements 
	in Fig.~\ref{Fig4}\textbf{f} coincide with those determined from the low-temperature 
	MCE and DMRG results. However, the V-state observed in neutron diffraction is much 
	narrower, which can be ascribed to the finite temperature of $95$~mK where our 
	diffraction measurements are conducted. As shown in Fig.~\ref{Fig3}\textbf{b}, the 
	diffraction scan fits excellently to the finite-$T$ phase diagram and only probes a 
	corner of the supersolid V phase.
	
	The onset temperatures of the Y, UUD, and V phases, can also be determined by 
	temperature-dependent neutron diffraction, which are found to be approximately 
	$154$~mK, $353$~mK, and $107$~mK (Fig.~\ref{Fig4}\textbf{e}), respectively,
	and in full agreement with the specific heat peaks (Fig.~\ref{Fig3}\textbf{b}). More 
	importantly, at $95$~mK, the magnetic diffraction intensities in both the Y and V 
	phases are significantly weaker compared to that of the UUD state, indicating a
	much smaller ordered moment. Through a preliminary magnetic structure refinement, 
		a maximal moment is estimated to be approximately $0.59$ $\mu\rm_{B}$ and $1.74$ 
		$\mu\rm_{B}$ is estimated for $B = 0$ and $0.8$~T, respectively (Methods). These 
	findings reveal the coexistence of magnetic ordering and strong fluctuations in the spin 
	supersolid phases.
	\\
	
	\noindent{\bf Discussion}\\  
	Nearly two decades after the theoretical proposal~\cite{Melko2005,Wessel2005,
		Heidarian2005,Prokofev2005}, here we have identified experimental signatures 
	of supersolidity in the triangular-lattice magnet NBCP by mutually corroborative 
	magnetothermal and neutron measurements, further supported by theoretical 
	calculations. Echoing the rich diversity of quantum spin states proposed in 
	the TLAF systems~\cite{Heidarian2010,Yamamoto2015,Sellmann2015}, 
	an increasing number of triangular lattice compounds emerge due to the 
	rapid experimental progresses. They share similar structures (Methods) 
	and may also host the exotic states like the spin supersolid. Furthermore, 
	the concept of spin supersolid has also been proposed in other systems, 
	such as spin-$1$ chains~\cite{Sengupta2007chain} and Shastry-Sutherland 
	lattice systems~\cite{Mila2008SS}, which remain to be experimentally 
	investigated.
	
	The MCE response in NBCP is much larger than in other magnetic materials 
	studied to date. The frustrated quantum magnets not only provide a versatile 
	and powerful platform for exploring novel spin states, they also possess 
	several attractive merits --- strong spin fluctuations enhanced by quantum 
	criticality and frustration effects, higher density of magnetic ions, and good 
	chemical stability --- as compared to hydrate paramagnetic coolants. 
	These advantages open up a promising avenue for their applications in 
	sub-Kelvin refrigeration for quantum technologies and space applications.
	\\
	%  ====== Bib ======= %
	% \bibliography{MCERef}

	\newpage
	\noindent {\bf Methods}
	\\
	\noindent 
	{\bf Samples}\\ 
	The polycrystalline samples of NBCP were firstly synthesized by the solid-state 
	reaction method, as reported in Ref.~\cite{Zhong2019}. Dried Na$_2$CO$_3$ 
	($99.99$\%), BaCO$_3$ ($99.95$\%), CoO ($99$\%) and (NH$_4$)$_2$HPO$_4$ 
	($99.99$\%) were mixed stoichiometrically and well ground, together with the 
	catalyst NH$_4$Cl ($99.99$\%) in a molar ratio of $2:1$. 
	The mixture was pressed into pellets, placed in an alumina crucible, and sintered 
	in air at $800~^\circ$C for $24$ hours. This sintering process was repeated for 
	several times to minimize possible impurities. After that, the single-crystal samples 
	of NBCP were grown using the flux method. The pro-obtained polycrystalline NBCP 
	with high purity and the NaCl ($99.99$\%) flux were mixed with a molar ratio of 
	$1:5$ and thoroughly ground. The mixture was transferred into a Pt crucible, 
	heated up to $950~^\circ$C for $2$ hours, and then cooled to $750~^\circ$C with 
	a rate of $3~^\circ$C/h. After cooling down to the room temperature in the end, 
	millimeter-sized single crystals of NBCP were obtained.
	
	The crystal structure of NBCP was characterized by a Bruker D$8$ ADVANCE
	diffractometer, with the diffraction patterns shown in Extended Data 
	Figs.~\ref{EDFig1}\textbf{a} and \textbf{b}, confirming the high purity and 
	good quality of the polycrystalline and single-crystal samples, respectively.  
	The single-crystal samples used for neutron diffraction experiments have 
	been checked and oriented by X-ray Laue photography, as shown in 
	Extended Data Figs.~\ref{EDFig1}\textbf{c} and \textbf{d}. Both crystals 
	are mounted with GE varnish onto a Cu plate fixed inside a Cu screw, 
	with the ($H$, $H$, $L$) reciprocal plane lying horizontally.
	\\
	
	\noindent
	{\bf Spin-$1/2$ model and boson mapping}\\ 
	As shown in Extended Data Fig.~\ref{EDFig2}, the lowest-energy 
	electron structure of a free Co$^{2+}$ ion is $^4F$ ($L = 3, S = 3/2$), 
	which splits under the regular octahedral crystal electric field (CEF) 
	and the ground state is $12$-fold degenerate (denoted as $^4T_{1g}$). 
	The effect of spin-orbit coupling further splits the degenerate energy 
	levels, and the two lowest levels of the high-spin Co$^{2+}$ ions in 
	NBCP form effective $S=1/2$. 
	
	The triangular lattice antiferromagnetic (TLAF) model for NBCP contains 
	the nearest-neighbor (NN) Heisenberg exchange, pseudo dipolar coupling 
	$J_{\rm PD}$, and the off-diagonal symmetric term $J_{\Gamma}$. 
	The form of the spin Hamiltonian is dictated by the $P\overline{3}m1$
	space group and particularly the site symmetry. To be concrete, 
	the coupling between two spins $S_i^\alpha$ and $S_j^\beta$ 
	($\alpha,\beta = x, y, z$) can be written as 
	$$H_{ij} =\sum_{\alpha,\beta} J^{\alpha \beta} S_i^{\alpha} S_j^{\beta}$$ 
	on the NN bond $\langle i,j \rangle$, with the coupling constants
	\begin{equation}
		J^{\alpha\beta} = 
		\begin{pmatrix}
			J_{xy}+2 J_{\rm PD} \cos\varphi
			&-2J_{\rm PD} \sin\varphi
			&-J_{\Gamma} \sin\varphi \\
			-2J_{\rm PD} \sin\varphi
			&J_{xy}-2J_{\rm PD} \cos\varphi
			& J_{\Gamma} \cos\varphi\\
			-J_{\Gamma} \sin\varphi
			& J_{\Gamma} \cos\varphi
			& J_z
		\end{pmatrix}. \nonumber
	\end{equation}
	The angle $\varphi = 0$, $2\pi/3$, and $4\pi/3$ for three different bonds. 
	To reveal the spin-spin interactions in NBCP, it resorts to determining the 
	parameters in the microscopic spin model. 
	
	A systematic parameter searching approach is by thermal data analysis
	\cite{Yu2021}, which finds that the anisotropic couplings $J_{\rm PD}$ 
	and $J_{\Gamma}$ are negligibly small ($\lesssim 50$~mK), and the 
	discrete angle $\varphi$ is irrelevant~\cite{Gao2022QMats}. The compound 
	can thus be accurately described by the easy-axis TLAF model with spin 
	rotational U($1$) symmetry, i.e.,
	\begin{equation}
		\label{Eq:TLAF}
		H_{ij} =  J_{xy} (S_i^x S_j^x +S_i^y S_j^y) 
		+ J_z S_i^z S_j^z - B \sum_i g_c \mu_B  S_i^z, 
	\end{equation}
	with $J_{xy} = 0.88$~K, $J_z = 1.48$~K, the Land\'e factor $g_c \simeq 4.89$
	\cite{Gao2022QMats}. Lately, a neutron scattering study of NBCP also determines
	the Hamiltonian as an easy-axis Heisenberg model with $J_{xy} = 0.882$($7$)~K, 
	$J_z = 1.451$($9$)~K by fitting the spin-wave dispersions, showing excellent agreement.
	
	The above spin-$1/2$ states can be mapped into hardcore bosons, 
	where the spin operators can be expressed as boson creation 
	($b^{\dagger}$) and annihilation ($b$) operators, i.e.,
	\begin{equation}
		S_i^+ \rightarrow b_i^\dagger,~S_i^- \rightarrow b_i,~S_i^z \rightarrow n_i-1/2, \notag
	\end{equation}
	where $n_i=b_i^\dagger b_i$ is the boson number operator. They
	satisfy the commutation relation $[b_i, b_j^\dagger] = \delta_{ij}(2 n_i -1)$.
	With this exact mapping, the TLAF model in Eq.~(\ref{Eq:TLAF}) can 
	be represented as a hardcore boson model
	\begin{equation}
		H =  -t \sum_{\langle i,j \rangle} (b_i^\dagger b_j + h.c.) 
		+ V \sum_{\langle i,j \rangle}  n_i n_j -\sum_i \mu \, n_i,
	\end{equation}
	with $t = -J_{xy}/2, V = J_z$, and $\mu =J_z + B$. 
	
	With this exact spin-boson mapping, the diagonal charge density wave order 
	in the boson system characterized by the structure factor $S_{nn}(q) = 
	\frac{1}{N^2} \sum_{i,j} \langle (n_i-1/2) (n_j-1/2) \rangle \, e^{i q \cdot {r}_{ij}}$ 
	and the off-diagonal superfluid order by $S_{\rm SF}(q) = \frac{1}{N^2} 
	\sum_{i\neq j} \langle b_i^\dagger b_j + h.c. \rangle \, e^{i q \cdot {r}_{ij}}$ 
	can be computed equivalently via the longitudinal spin structure 
	factor $S_{zz}(q) = \frac{1}{N^2} \sum_{i,j} \langle S_i^z S_j^z \rangle 
	\, e^{i q \cdot {r}_{ij}}$ and transverse components $S_{xy}(q) = \frac{1}{N^2} 
	\sum_{i\neq j} \langle S_i^+ S_j^- + h.c. \rangle \,e^{iq \cdot {r}_{ij}}$, 
	respectively. The results of solid and superfluid order parameters shown 
	in Fig.~\ref{Fig4}\textbf{g} are calculated with \emph{\textbf{q}} $  \equiv {\rm K} = (1/3, 1/3)$.
	\\
	
	\noindent 
	{\bf Tensor network simulations} \\
	We exploit the exponential tensor renormalization group (XTRG) method
	\cite{Chen2018,Lih2019,Chen2018b} to simulate the finite-temperature 
	properties of the easy-axis TLAF model for NBCP. The XTRG calculations 
	start from the initial density matrix $\rho_0(\tau)$ at very high temperature 
	$T \equiv 1/\tau$ with $\tau \ll 1$, represented in a matrix product operator 
	(MPO) form~\cite{ChenSETTN2017}. The series of density matrices 
	$\rho_n(2^n \tau)$ at various lower temperatures are obtained by iteratively 
	squaring the density matrix $\rho_n = \rho_{n-1} \cdot \rho_{n-1}$ to cool 
	down the system exponentially. After each MPO multiplication, truncations 
	of the MPO bond bases are conducted to make the cooling process sustainable.
	
	In practice, we perform the XTRG calculations on cylinders with width 
	$W = 6$ and length up to $L = 15$ (on Y-type cylinder with one edge 
	parallel to the circumference direction), and retain up to $D = 2,000$ 
	bond states that guarantee a high accuracy down to low temperatures 
	(below $100$~mK). The partition function at inverse temperature $\beta 
	\equiv 1/T$ can be obtained as $\mathcal{Z}(\beta) =  {\rm tr} [\rho(\beta/2) 
	\rho(\beta/2)^\dagger]$ via the bilayer trace~\cite{Dong2017,Chen2018,
		tanTRG2023}, based on which the thermodynamics quantities like 
	$S_{\rm m}$ and $C_{\rm m}$ can be obtained.
	
	Besides the finite-$T$ properties, we also employ the density matrix 
	renormalization group (DMRG) method to accurately simulate the 
	ground-state properties (i.e., $T=0$), including the spin-spin correlations 
	and corresponding structure factors (see Fig.~\ref{Fig4}\textbf{g}). 
	In practice, we compute the easy-axis TLAF model by retaining $D=2,000$ 
	bond states, which produce accurate and well converged results with small 
	truncation errors $\epsilon \lesssim 1 \times 10^{-5}$. Moreover, in practical 
	DMRG calculations we compute the results on YC$6\times9$ (width $W=6$ 
	and length $L=9$) and YC$6\times15$ ($L=15$) lattices, and calculate the 
	bulk properties by subtracting the results on the two geometries to further 
	reduce finite-size effects. 
	\\
	
	\noindent 
	{\bf Specific-heat measurements and magnetic entropy}\\ 
	The specific heat measurements were conducted with a quasi-adiabatic 
	heat pulse method in the $^3$He–$^4$He DR insert with a single-crystal 
	sample of $0.85$~mg, under magnetic fields $B$ // $c$ axis and within 
	the temperature range $0.06$~K $\le T \le 12$~K.  
	
	In Extended Data Fig.~\ref{EDFig4}\textbf{a}, we fit the low-temperature 
	specific heat data under two fields ($0.75$~T and $2$~T) with a sum of nuclear 
	Schottky term ($\alpha_{\rm n} \cdot T^{-2}$) and electron spin contribution 
	$\beta_{\rm e} \, T^{-1} e^{-\Delta/T}$, with $\Delta$ the spin excitation gap. 
	The corresponding fitting results are shown in Extended Data 
	Figs.~\ref{EDFig4}\textbf{c, d}. 
	For the cases with different fields, especially those within the supersolid 
	regime where spin excitations become gapless and thus the electron spin 
	contributions overlap strongly with the nuclear part, we assume a linear 
	relation $\alpha_{\rm n} \propto B$ (Extended Data Fig.~\ref{EDFig4}\textbf{b}), 
	which is widely adopted in literatures~\cite{NHC2013}. With the interpolated 
	and extrapolated $\alpha_n$ coefficients, we compute the nuclear spin 
	contributions, as shown in Extended Data Figs.~\ref{EDFig4}\textbf{e-g}. 
	By subtracting the fitted nuclear part and the phonon contributions (estimated 
	from the nonmagnetic isostructural compound Na$_2$BaMg(PO$_4$)$_2$
	\cite{Zhong2019}) from the measured total specific heat $C_{\rm p}$, we thus 
	obtain the magnetic specific heat $C_{\rm m}$. 
	
	With the specific heat data in Extended Data Figs.~\ref{EDFig4}, we compute 
	the magnetic entropy $$S_{\rm m}(B) = R \, {\rm ln}2 - \int_{T_{\rm max}}^{T_{\rm min}}
	\frac{C_{\rm m}(T, B)}{T}dT$$ and show the results in Extended Data Figs.~\ref{EDFig4}\textbf{h-j}. 
	The $T_{\rm min} \simeq 50$-$70$~mK is the lowest temperature in the 
	heat capacity measurement, and $T_{\rm max} \simeq 10$-$12$~K is the 
	upper bound of integration. Based on the experimental and calculated 
	entropy results, we estimate the cooling temperatures and validate that 
	the giant magnetic cooling effect observed in NBCP is ascribed to the 
	electron spin fluctuations, while the nuclear contribution does not play 
	any essential role here.
	\\
	
	\noindent 
	{\bf Magnetocaloric measurements} \\ 
	In the magnetocaloric effect (MCE) measurements starting from a 
	relatively high initial temperature ($T_0 \geq 2$~K), single crystals 
	with a total weight of $1.65$~g are fixed on the sliver stage by cryogenic 
	GE varnish and formed into a column-like shape (see Extended Data 
	Fig.~\ref{EDFig3}\textbf{a}) with the $c$ axis oriented along the 
	magnetic field. In practical measurements, the sample column is cooled 
	down to $2$~K from $40$~K. When the sample column is stabilized 
	at $2$~K, we pump the residual gas in the chamber to obtain adiabatic 
	conditions with a cryopump. Demagnetization cooling measurements 
	are performed by gradually decreasing fields from the initial value 
	$B_0 = 3$-$4$~T at a rate of $dB/dt = 0.6$~T$\cdot$min$^{-1}$.
	
	For low initial temperatures below $300$~mK, a standard calorimetric
	puck installed to the $^3$He-$^4$He DR insert of PPMS is used. 
	The thermometer has been carefully calibrated at various temperatures 
	($50$~mK-$4$~K) and under a wide range of magnetic fields ($0$-$4$~T). 
	The polymer strips are used for mechanical support of the sample platform. 
	Due to the strong magnetic anisotropy of the compound, we adopt a small
	NBCP sample with a mass of $3.2$~mg, to avoid large magnetic torque 
	that may break the polymer strips. In practical DR-based measurements, 
	the field ramps down from $B_0 = 3$~T with a speed of $0.03$-$0.12$~T$\cdot$min$^{-1}$. 
	
	We compare the magnetic cooling performances between NBCP 
	and the commercial refrigerants Gd$_3$Ga$_5$O$_{12}$ (GGG), 
	Fe(SO$_4$)$_2$ (NH$_4$)$_2$ $\cdot$ 6H$_2$O (FAA), and 
	CrK(SO$_4$)$_2\cdot12$H$_2$O (CPA) widely used in the 
	sub-Kelvin applications. The setup for GGG sample measurements 
	are shown in Extended Data Fig.~\ref{EDFig3}\textbf{b}.
	As the GGG sample mass is as large as $50$~g, the parasitic heat 
	leakage has thus limited impact in its demagnetization cooling. As 
	shown in Extended Data Fig.~\ref{EDFig6}\textbf{a}, although 
	the NBCP single crystal is with much smaller mass of only $1.65$~g, 
	it nevertheless reaches significantly lower temperatures than GGG, 
	given the same initial conditions $T_0 = 2$-$4$~K and $B_0 = 4$~T.
	
	Representative paramagnetic salts FAA and CPA are measured with 
	the same setup as for NBCP (Extended Data Fig.~\ref{EDFig3}\textbf{a}). 
	As space/volume of uniform field generated by the superconductor 
	magnet is usually limited, and volumetric entropy density is important in 
	practical refrigeration applications, we use single-crystal 
	samples with similar volumes (about $0.4$~cm$^3$) to make a fair
	comparison. In Extended Data Fig.~\ref{EDFig6}\textbf{b},
	we find NBCP clearly outperforms the two commercial paramagnetic 
	refrigerants.
	
	In Extended Data Figs.~\ref{EDFig6}\textbf{c, d}, we compare the cooling 
	efficiency, capacity, and hold time between NBCP and two hydrate salts. 
	The efficiency $\eta$ is the ratio of heat absorption $\Delta Q_{\rm c}$ 
	(the purple hatched region in Extended Data Fig.~\ref{EDFig6}\textbf{c}) 
	and the heat release $\Delta Q_{\rm m}$ during the magnetization process 
	at the environment temperature $T_0$ (grey shaded region). We find a high 
	efficiency of $\eta \simeq$ $23\%$ for NBCP, much higher than those 
	of paramagnetic coolants FAA ($9\%$) and CPA ($11\%$) (data taken 
	from Ref.~\cite{Wolf2011}). Therefore, NBCP can deliver more cooling 
	power with less heat production in a magnetic refrigeration circulation. 
	
	The high density of magnetic ions in NBCP offers a large magnetic 
	entropy density $S^V_{\rm m} =56$~mJ $\cdot$ K$^{-1}\cdot$cm$^{-3}$, 
	which is even slightly higher than those of higher-spin paramagnetic 
	coolants, e.g., $S = 5/2$ FAA ($53$~mJ $\cdot$K$^{-1}\cdot$cm$^{-3}$) 
	and $S = 3/2$ CPA ($42$~mJ $\cdot$K$^{-1}\cdot$cm$^{-3}$). Such 
	a high entropy density leads to great cooling capacity per volume for 
	sub-Kelvin refrigeration. The large heat absorption $\Delta Q_{\rm c}$ 
	of NBCP further emphasizes this aspect. In Extended Data 
	Fig.~\ref{EDFig6}\textbf{d}, we see that this is indeed the case, and 
	NBCP remains below $1$~K for $1.5$~h, much longer than those of 
	FAA and CPA under the same condition.
	
	Based the above comparisons, and considering that NBCP has good 
	chemical stability in high vacuum --- an important advantage over hydrate 
	salts for practical sub-Kelvin applications, we conclude that NBCP is a 
	promising quantum magnetic coolant with exotic supersolid spin states 
	and superior refrigeration performance.
	
	Moreover, in recent years there has been a surge of interest in investigating 
	other triangular-lattice magnets~\cite{Zhou2012,Yutaka2012PRL,BCSO2015PRB,
		NYS2019PRB,YMGO2020}, including A$_2$BM(PO$_4$)$_2$ (A is an alkali 
	element and M=Ni, Mn, and B=Ba, Sr)~\cite{LiN2021NBNP,Kim2022NBMP,
		Zhang2022NBSCP}, A$_2$Co(SeO$_3$)$_2$~\cite{RCSO2020}, and rare-earth 
	triangular magnets A$_3$RE(PO$_4$)$_2$ (with RE being a rare-earth ion)
	\cite{Guo2020RYPO}, as well as Ba$_3$REB$_3$O$_9$/Ba$_3$REB$_9$O$_{18}$
	\cite{Cho2021BYBO,Khatua2022Ba3RB9O18}, ABaRE(BO$_3$)$_2$
	\cite{Guo2019NBYB,Tokiwa2021}. Some of these compounds also share the 
	same lattice structure and site symmetry, meaning that their spin-spin interactions 
	follow Eq.~(\ref{Eq:TLAF}) and may also host exotic quantum spin states and 
	serve as high-performance magnetic coolants.
	\\
	
	{
		\noindent 
		{\bf Single-crystal neutron diffraction} \\ 
		Single-crystal neutron diffraction experiments were conducted on the 
		high-intensity neutron diffractometer WOMBAT ~\cite{Wombat} at Australia’s 
		Nuclear Science and Technology (ANSTO) and the thermal-neutron 
		two-axis diffractometer D$23$ at Institut Laue Langevin (Grenoble, France). 
		The incident neutron wavelength at WOMBAT and D$23$ is chosen as 
		$2.41$~\AA~ and $2.387$~\AA~, respectively. At both instruments, 
		the ($-1$, $1$, $0$) direction of the crystal was aligned vertically, so that 
		the ($H$, $H$, $L$) reciprocal-space plane can be mapped out by 
		rotating the sample. 
		
		At WOMBAT, no magnetic field was applied. A single-crystal sample 
		with a mass of $\sim50$~mg was glued on a copper plate and 
		put into a top-loading cryostat equipped with the dilution refrigeration 
		insert, allowing to reach a base temperature of $30$ mK. The 
		reciprocal-space scans across the ($1/3$, $1/3$, $0.183$) magnetic peak 
		are shown in Extended Data Figs.~\ref{EDFig8}\textbf{a, b}. By fitting 
		the peak profiles using a Gaussian function, the widths of this magnetic 
		peak are found to be almost limited by the instrumental resolutions, 
		which are approximated by the full-width at half-maximum (FWHM) of 
		the nuclear reflections, indicating a long-range ordered nature 
		of the spin supersolid state stabilized by interlayer couplings and 
		ruling out the possible quantum spin liquid state scenario.
		
		At D$23$, a much larger single-crystal sample of NBCP with a mass 
		of $\sim$ $125$~mg was glued on a copper plate and put inside a $3.8$~T 
		horizontal field cryomagnet equipped with the dilution refrigeration insert. 
		A magnetic field up to $2$ T along the $c$ axis were applied and a base 
		temperature of $\sim90$~mK was achieved. The magnetic reflections 
		were recorded as a function of temperature in different magnetic fields, 
		as shown in Extended Data Figs.~\ref{EDFig8}\textbf{c-e}.
		\\
		
		\noindent 
		{\bf Magnetic structure refinement} \\
		To quantitatively compare the ordered moment sizes in the spin supersolid 
		and spin solid states, we have tried the magnetic structure refinement 
		based on the integrated intensities of the magnetic reflections for $B = 0$ 
		and $B= 0.8$~T, respectively. Due to the blockage of the bulky horizontal 
		magnet, only $7$ non-equivalent incommensurate magnetic reflections for 
		$0$~T and $10$ non-equivalent commensurate magnetic reflections for $0.8$~T 
		constrained in the ($H$, $H$, $L$) plane are accessible. Therefore, 
		though we were not able to carry out a strict full magnetic structure 
		determination, a ``preliminary" refinement can be conducted to confirm 
		the main spin direction along $c$ and to estimate the ordered moment size. 
		
		According to the irreducible representation analysis performed using the 
		BASIREPS program integrated into the FULLPROF suite~\cite{Fullprof}, 
		for an incommensurate magnetic propagation vector of \emph{\textbf{k}} $ = $($1/3$, $1/3$, $0.166$), 
		the magnetic representation $\Gamma\rm_{mag}$ for the Co$^{2+}$ ions 
		can be decomposed as the sum of three irreducible representations (IRs), 
		$\Gamma_{\mathrm{mag}}=\Gamma_{1}^{1}\oplus\Gamma_{2}^{1}\oplus\Gamma_{3}^{1},$ 
		whose basis vectors are listed in Extended Data Fig.~\ref{EDFig9}\textbf{a}. 
		As shown in Extended Data Fig.~\ref{EDFig9}\textbf{c}, the integrated intensities 
		of the $7$ incommensurate magnetic reflections for $B = 0$ can be relatively 
		well fitted by a modulated UUD structure described by the IR $\Gamma_{1}$, 
		with the main moments pointing along the $c$ axis. The moment sizes on 
		the $z=0$ layer are estimated to be $0.593$($6$), $-0.297$($3$) and $-0.297$($3$) 
		$\mu\rm_{B}$, for the ``up'', ``down'' and ``down'' spins, respectively.
		The reason why a Y-like spin configuration cannot be directly determined here is 
		likely due to the small size of the ordered in-plane moments at $95$~mK, especially
		considering the strong in-plane fluctuations of the spin supersolid state. 
		
		In contrast, for a commensurate \emph{\textbf{k}} $=$($1/3$, $1/3$, $0$), the magnetic representation 
		$\Gamma\rm_{mag}$ for the Co$^{2+}$ ions can be decomposed as the sum of 
		two IRs, $\Gamma_{\mathrm{mag}}=\Gamma_{2}^{1}\oplus\Gamma_{3}^{2},$ 
		whose basis vectors are listed in Extended Data Fig.~\ref{EDFig9}\textbf{b}. 
		As shown in Extended Data Fig.~\ref{EDFig9}\textbf{d}, the integrated intensities 
		of the $10$ commensurate magnetic reflections for $B = 0.8$ T fit well with a 
		non-modulated UUD structure described by the IR $\Gamma_{2}$. The moment 
		sizes on each layer are estimated to be $1.74$($3$), $-0.87$($2$) and $-0.87$($2$) $\mu\rm_{B}$, 
		for the ``up'', ``down'' and ``down'' spins, respectively. Please note that the true UUD
		spin structure in the spin solid state is a ``two-$k$'' ferrimagnetic structure consisting 
		of an antiferromagnetic structure with \emph{\textbf{k}} $= $($1/3$, $1/3$, $0$) and a ferromagnetic structure 
		with $k = 0$. Our refinement here is only able to resolved the antiferromagnetic 
		contribution from the \emph{\textbf{k}} $= (1/3$, $1/3$, $0)$ part, and it is very hard to accurately measure 
		the ferromagnetic scattering intensities superimposed on top of the strong nuclear 
		reflections accessible in the ($H$, $H$, $L$) plane. 
		
		Polarized neutron diffraction experiments at very low temperature, although highly 
		challenging, would be very helpful to identify the small ordered in-plane moments 
		in the spin supersolid phase.
		\\
		
		\bibliography{./MCERef.bib} 
		
		$\,$\\
		\textbf{Acknowledgements} \\
		The authors are indebted to Dehong Yu, Long Zhang, Shang Gao, 
		Tao Shi, and Xue-Feng Zhang for helpful discussions, Eric Ressouche, 
		Ketty Beauvois and Pascal Fouilloux for technical supports at ILL, 
		and Xingye Lu and Yi Li for the help with the orientation of the 
		single-crystal sample using x-ray Laue diffraction. This work was 
		supported by the National Natural Science Foundation of China 
		(Grant Nos.~12222412, 12074023, 11974036, 12047503, 12074024, 
		11834014, 52088101, and 12141002), Strategic Priority 
		Research Program and {Scientific Instrument Developing Program} 
		of CAS (Grant Nos.~XDB28000000, {ZDKYYQ20210003}), CAS 
		Project for Young Scientists in Basic Research (Grant No.~YSBR-057), 
		and the Fundamental Research Funds for the Central Universities in China. 
		We thank the HPC-ITP for the technical support and generous allocation 
		of CPU time. This work was supported by the Synergetic Extreme 
		Condition User Facility (SECUF) and the beamline 1W1A of the 
		Beijing Synchrotron Radiation Facility. The Australian Center for 
		Neutron Scattering and Institut Laue-Langevin are gratefully acknowledged 
		for providing neutron beam time through proposal P14250 and 5-41-1193.
		
		$\,$\\
		\textbf{Author contributions} \\
		W.L., J.X., and G.S. initiated this work. C.Z., B.L., and W.J. prepared 
		the samples. J.X., and P.S. designed and performed the magnetocaloric 
		measurements. J.X., and G.L. conducted the low-temperature specific 
		heat measurements. J.X., Z.C., J.S., H.J., P.S., W.L., and G.S. 
		conducted the magnetocaloric and specific heat data analysis. 
		W.S., K.S., C.W., C.Z., and W.J. performed the neutron scattering 
		experiments. C.Z., Y.G., W.L., and W.J., analyzed the neutron data. 
		Y.G., N.X., X.Y.L., Y.Q., Y.W., and W.L. conducted the microscopic 
		spin model analysis and performed the many-body calculations. 
		W.L., J.X., W.J., Y.G., P.S., and G.S. wrote the manuscript with input
		from all coauthors. W.J., W.L., P.S., and G.S. supervised the project. 
		
		$\,$\\
		\textbf{Competing interests} \\
		The authors declare no competing interests. 
		
		$\,$\\
		\textbf{Data availability} \\
		The data that support the findings of this study are available at 
		\href{https://www.nature.com/articles/s41586-023-06885-w}{https://www.nature.com/articles/s41586-023-06885-w} and from the corresponding 
		author upon reasonable request.
		
		$\,$\\
		\textbf{Code availability} \\
		The code that supports the findings of this study is available 
		from the corresponding author upon reasonable request.
		
		$\,$\\
		\textbf{Additional information} \\
		\textbf{Supplementary Information} is available in the online version of the paper. \\
		\textbf{Correspondence and requests for materials} should be addressed to 
		Wentao Jin, Wei Li, Peijie Sun, or Gang Su.
		
		% ======================
		% Extended figures
		% ======================
		$\,$\\
		\textbf{Extended Data} \\ 
		\setcounter{figure}{0}  
		\renewcommand{\figurename}{\textbf{Extended Data Figure}}
		
		% =============== Extended Data Figures =============== %
		\begin{figure*}[t]
			\includegraphics[width=1\textwidth]{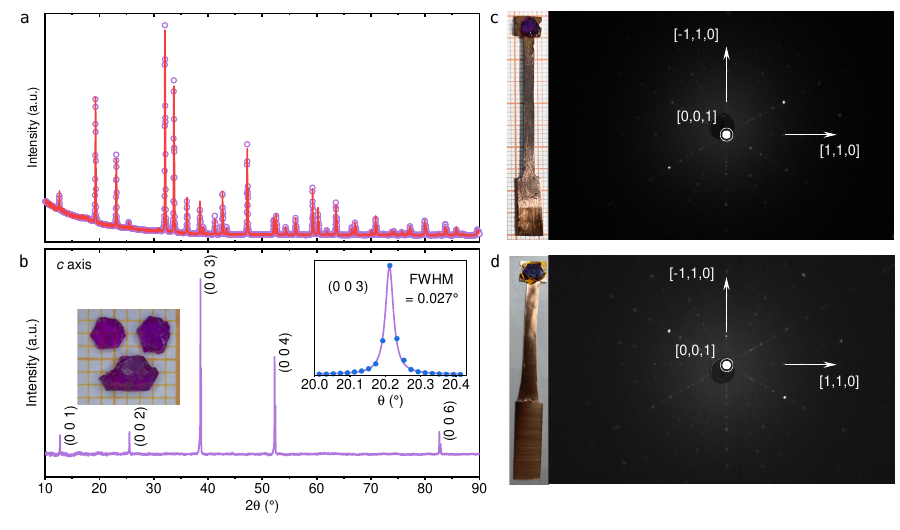}
			\caption{\textbar~\textbf{Sample structure characterization.} 
				\textbf{a} Powder X-ray diffraction pattern of NBCP measured 
				at room temperature. The red line indicates the calculated diffraction pattern 
				for comparison. 
				\textbf{b} Single-crystal X-ray diffraction pattern recorded for the ($0$,$0$,$L$) 
				planes at room temperature. The insets show the photo of the NBCP 
				single crystals grown from the same batch, which are used for the specific 
				heat, magnetocaloric and neutron measurements, and the rocking-curve scan of the 
				($0 0 3$) reflection for one representative crystal. The very sharp peak profile 
				with a full width at half maximum (FWHM) of $0.027^{\circ}$ indicates the high 
				quality of the crystals. 
				\textbf{c, d} Images and X-ray Laue patterns of the single-crystal samples 
				used for the neutron diffraction experiments at WOMBAT ($50$~mg, \textbf{c}) 
				and D$23$ ($125$~mg, \textbf{d}), respectively, mounted on the Cu plates. 
				Sharp and bright diffraction spots confirm good quality of the crystals.}
			\label{EDFig1}  
		\end{figure*}

		\begin{figure*}[]
			\includegraphics[width=0.7\linewidth]{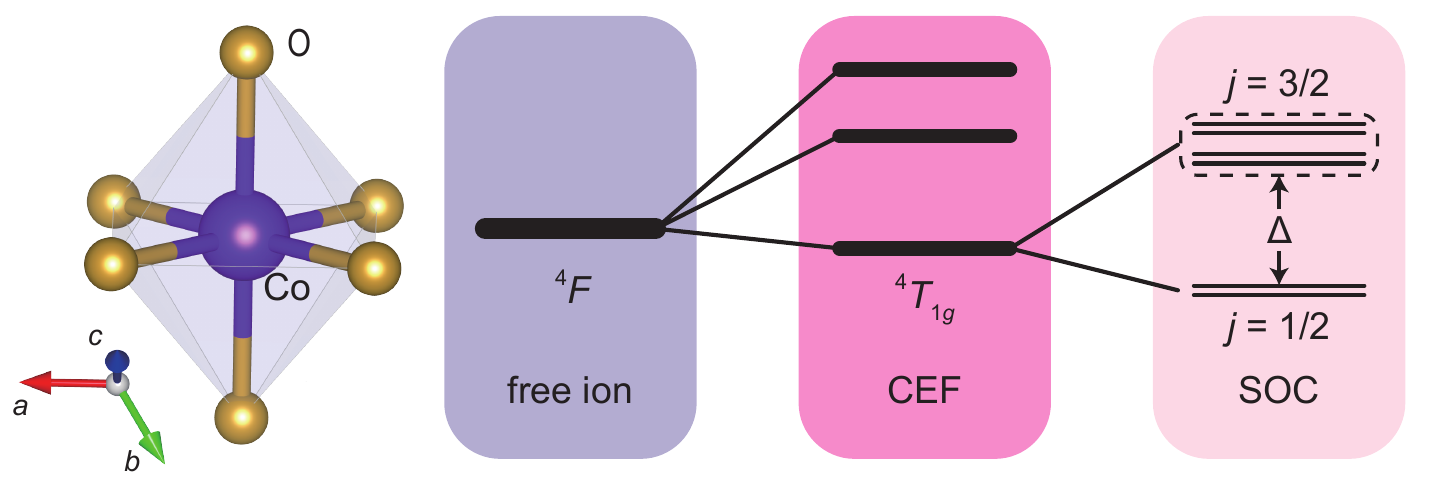}
			\caption{\textbar~{\bf Energy level diagram of the Co$^{2+}$ $^4F$ state.} 
				Under the effects of octahedral crystal electric field and spin-orbit coupling, 
				the lowest lying ground state is a $j=1/2$ Kramers doublet separated from 
				the first excited $j=3/2$ level by a gap of about $\Delta \approx 70$~meV
				\cite{Gao2022QMats}.
			}
			\label{EDFig2}  
		\end{figure*}

		% ========== Measurement Devices ========= %
		\begin{figure*}[t]
			\includegraphics[width=1\textwidth]{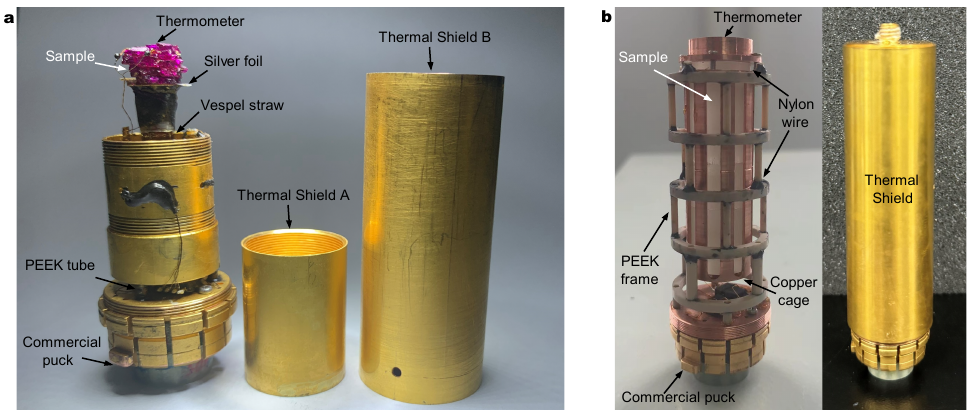}
			\caption{$\vert$
				\textbf{Photos of PPMS-based quasi-adiabatic demagnetization 
					cooling measurement setups.}
				\textbf{a} Setup for the measurements of the NBCP single-crystal 
				samples and the paramagnetic coolants FAA and CPA, with two
				gold-plated thermal shields. To enhance thermal insulation, we 
				employed a Vespel straw to support the sliver stage. We connected 
				the field-calibrated RuO$_2$ thermometer using two pairs of twisted 
				manganin wires ($25~\mu$m in diameter and approximately $30$~cm 
				in length) to minimize heat leakage. Additionally, two thermal shields
				were used to protect the sample column from thermal radiation and 
				other parasitic heat loads from the PPMS chamber. These shields 
				were connected using PEEK tubes. In Fig.~\ref{Fig2}\textbf{b}, we 
				show the setup for a relatively large mass ($50$~g) GGG sample 
				with a gold-plated thermal shield.}	
			\label{EDFig3}  
		\end{figure*}
		
		% ========== Specific heat ========= %
		\begin{figure*}[t]
			\includegraphics[width=1\textwidth]{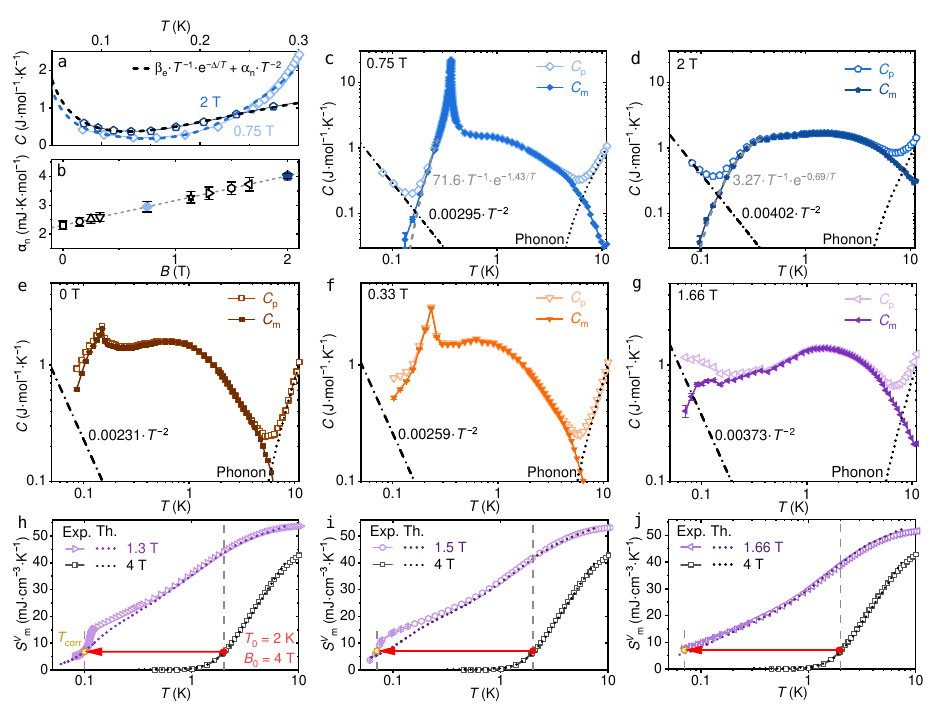}
			\caption{$\vert$~
				\textbf{Analysis of the specific heat and magnetic entropy results
					measured under magnetic fields.}
				\textbf{a} The measured specific heat data ($C_{\rm p}$) of NBCP 
				under magnetic fields of $0.75$~T and $2$~T, both of which turn 
				up at low temperature $T \lesssim 150$~mK due to the nuclear 
				spin Schottky anomaly. The open symbols are estimated parameter 
				$\alpha_n$ of nuclear spin contributions for different fields, {based
					on the fitted $\alpha_n$ under $0.75$~T and $2$~T, and}
				supposed to be linear vs. field $B$. The provided error bars are
				estimated based on the fitting uncertainty in the $B=0.75$~T case.
				\textbf{c-g} show the total specific heat $C_{\rm p}$ (open symbols)
				under different fields, which contains the fitted nuclear spin contribution 
				(dotted-dashed line) and phonon specific heat (dotted line). 
				After subtracting these contributions, we show the resulting 
				magnetic specific heat $C_m$ with solid symbols.
				\textbf{h-j} The magnetic entropy can be obtained by taking the 
				integration of the specific heat, and both measured data (Exp.) 
				and model calculation (Th.) are shown. 
				The red arrows indicate an ideal adiabatic demagnetization cooling 
				process, which starts from the initial conditions $T_0=2$~K and $B_0=4$~T 
				and end with lowest temperature $T_{\rm corr}$ shown in Fig.~\ref{Fig2}\textbf{a} 
				of the main text. The estimated error bars, which are based on the 
				fitting uncertainty of nuclear contribution, are smaller than the size of symbols.
			}
			\label{EDFig4}  
		\end{figure*}

		% ========== Magnon dispersion ========= %
		\begin{figure*}[htp]
			\includegraphics[width=1\linewidth]{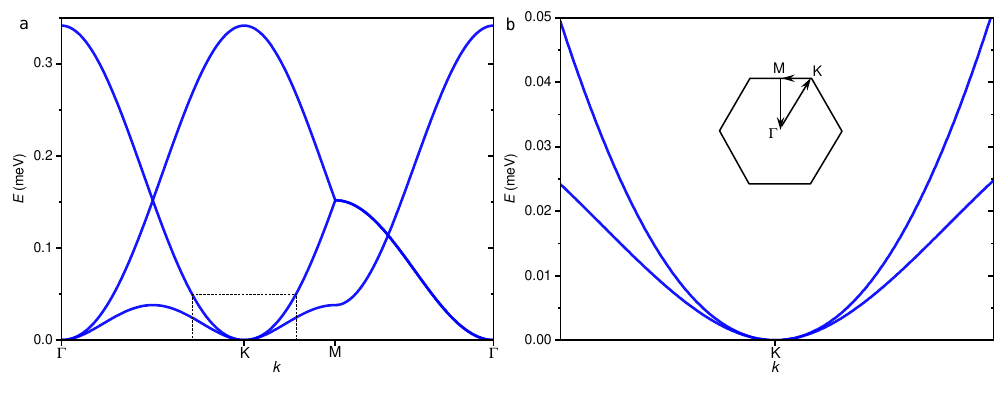}
			\centering
			\caption{$\vert$
				\textbf{The magnon dispersion by the linear spin-wave theory.}
				{\textbf{a} Magnon dispersion $E_k$ calculated by the linear spin-wave theory 
					at the supersolid transition $B_{c3}^*\simeq 1.69$($6$)~T, and the dashed box in 
					\textbf{a} is expanded in \textbf{b}. The easy-axis TLAF model simulated here 
					use the effective parameters, i.e., $J_{z}/J_{xy} = 1.68$, with two gapless quadratic 
					modes at the $\rm K$ point. {They enhance} the low-energy density of states 
					and give rise to significant entropic effect at low temperature. The results indicate 
					a quantum critical point (QCP) with dynamical exponent $z=2$. Since two order 
					parameters, the solid order associated with lattice rotational $\mathbb{Z}_3$ 
					symmetry breaking and the superfluid order with U($1$) spin symmetry {breaking}, 
					both vanish at the same quantum phase transition point (c.f., Fig.~\ref{Fig3}\textbf{b} in 
					the main text), we argue that there may exist an emergent $O$($4$) symmetry at the 
					QCP that include the $\mathbb{Z}_3 \times$ U($1$) as its subgroup. 
			}}
			\label{EDFig5}
		\end{figure*}

		% ========== Cooling performance ========= %
		\begin{figure*}[t]
			\includegraphics[width=0.9\textwidth]{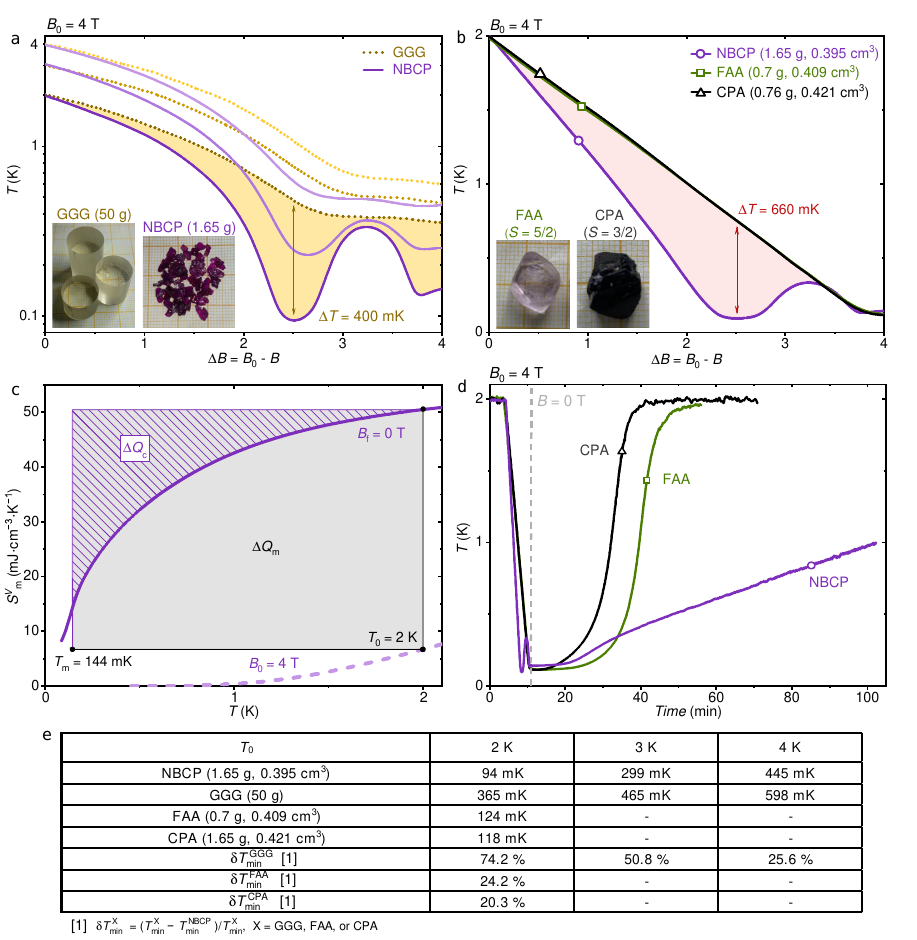}
			\caption{$\vert$
				\textbf{Comparisons on cooling performances.}
				\textbf{a} Quasi-adiabatic demagnetization cooling curves of NBCP 
				(solid line) and GGG (dashed line) from the initial field of $B_0=4$~T 
				and different base temperatures $T_0=2$-$4$~K. \textbf{b} compares 
				the demagnetization cooling of NBCP with paramagnetic coolants CPA 
				and FAA. The volumes of different materials are roughly equal (about 
				$0.4$~cm$^3$). NBCP cools down much faster in the high-field regime, 
				i.e., $B_{\rm c2}^* \leq B \leq B_{\rm c3}^*$ (V-type spin supersolid phase). 
				We find NBCP cools down to $94$~mK at about $1.5$~T, much lower than 
				that of paramagnets by $\Delta T\simeq 660$~mK. The insets in \textbf{a} 
				and \textbf{b} show single-crystal samples of NBCP ($1.65$~g, $0.395$
				cm$^3$), GGG ($50$~g), FAA ($0.7$~g, {$0.409$~cm$^3$}), and CPA 
				($0.76$~g, $0.421$~cm$^3$) used in practical MCE measurements.
				\textbf{c} shows the volumetric magnetic entropy densities of NBCP 
				at $B_{\rm f} = 0$~T (solid line) and $B_0 = 4$~T (dashed line). The 
				heat absorption of NBCP $\Delta Q_{\rm c} \simeq 19$~mJ$\cdot$cm$^{-3}$ 
				at zero field (indicated by the purple shaded area), which turns out to 
				be much higher than those of FAA ($3.0$~mJ$\cdot$cm$^{-3}$) and 
				CPA ($3.1$~mJ$\cdot$cm$^{-3}$) as reported in Ref.~\cite{Wolf2011}. 
				Therefore, NBCP is able to maintain the reached low temperatures for 
				a longer period of time (hold time). As shown in \textbf{d}, the NBCP 
				remains below $1$~K for about $1.5$ hours, while the temperatures 
				of CPA and FAA quickly rise up and reach thermal equilibrium with $2$~K 
				environment within $30$-$40$ mins. \textbf{e} tabulates the lowest cooling 
				temperatures in the quasi-adiabatic demagnetization process from the 
				same initial field $B_0 = 4$~T and various initial temperature $T_0$.
			}
			\label{EDFig6}  
		\end{figure*}

		% ========== Low temperature MCE ========= %
		\begin{figure*}[htp]
			\includegraphics[width=1\textwidth]{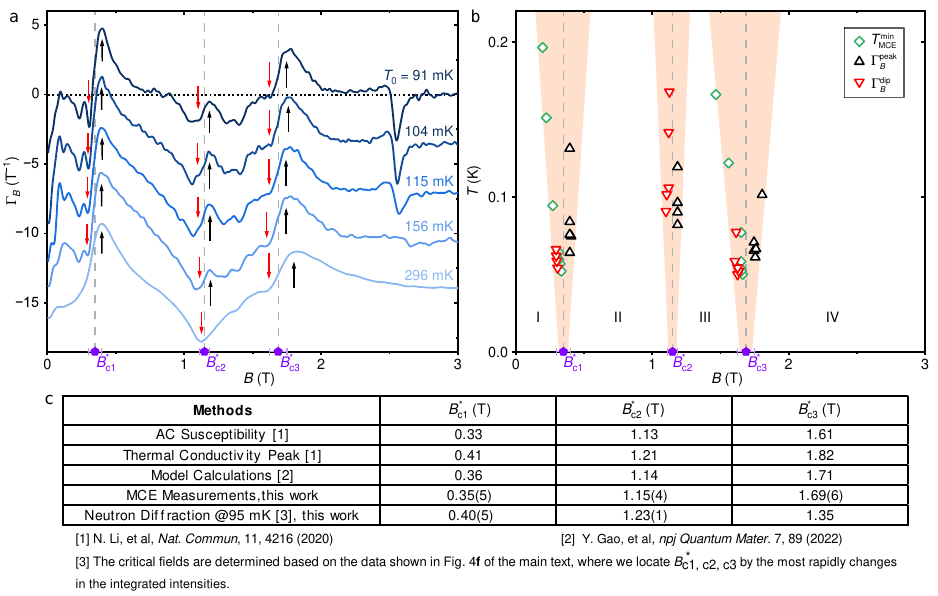}
			\caption{$\vert$ \textbf{Low temperature magnetic Gr\"uneisen ratio
					and location of quantum critical points.} \textbf{a} Magnetic Gr\"uneisen 
				ratio $\Gamma_B$ deduced from the isentropic $T$-$B$ lines, 
				$\Gamma_B = \frac{1}{T} (\frac{\partial T}{\partial B})_S$, with the red
				(black) arrows indicating the dip~(peak) locations. The quasi-adiabatic 
				cooling starting from various low initial temperatures $T_0 \leq 300$~mK, 
				and an initial field of $B_0=3$~T. The lines have been shifted with a 
				constant offset of $3.5$~T$^{-1}$ for the sake of clarity. 
				\textbf{b} Characteristic ($T$, $B$) values, including those determined 
				from $T^{\rm min}_{\rm MCE}$ (lowest temperature) and the peak 
				($\Gamma_B^{\rm peak}$) and dip ($\Gamma_B^{\rm dip}$) of 
				$\Gamma_B$. They naturally form quantum critical ``fans'' (orange 
				shaded areas) that converge to the QCPs ($B_{\rm c1}^*, B_{\rm c2}^*, 
				B_{\rm c3}^*$) in the low temperature limit, denoted by the purple 
				pentagons along with error bars estimated. \textbf{c} lists the field-induced 
				quantum phase transitions ($B$ // $c$ axis) in NBCP, and compare 
				our results to the previous experimental and theoretical works, where 
				excellent agreements are seen.}
			\label{EDFig7}  
		\end{figure*}
		
		% ========== Single-crystal neutron diffraction scan ========= %
		\begin{figure*}[htp]
			\includegraphics[width=1\textwidth]{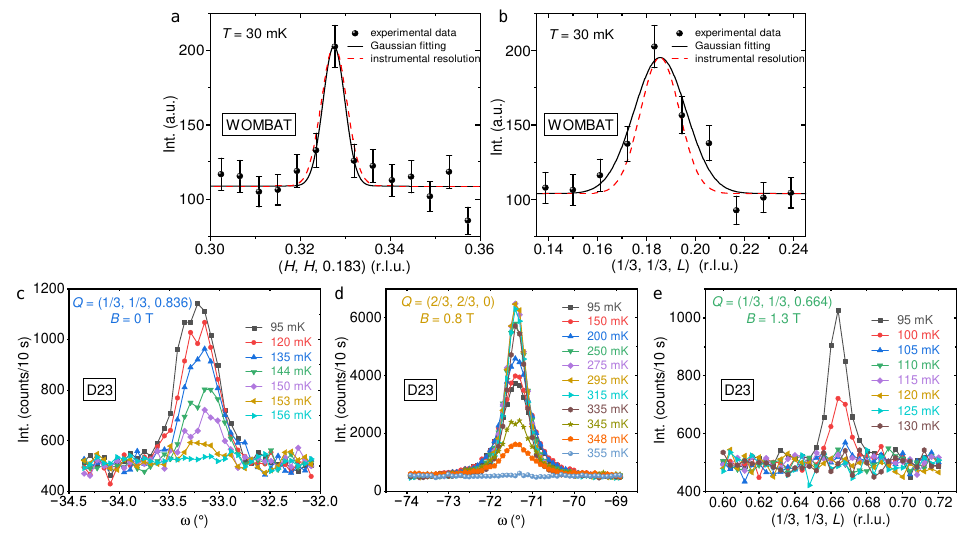}
			\caption{$\vert$
				\textbf{Single-crystal neutron diffraction scans.}
				Reciprocal-space scans at $30$ mK performed at WOMBAT in zero-field across 
				($1/3$, $1/3$, $0.183$), the strongest magnetic reflection, along \textbf{a} the 
				in-plane ($H$, $H$, $0.183$) direction and \textbf{b} out-of-plane ($1/3$, $1/3$, $L$) 
				direction, respectively. The solid and dashed lines represent the Gaussian 
				fittings to the experimental data and the instrumental resolutions represented 
				by the nuclear reflections, respectively. The fitted peak position in \textbf{a} is 
				slightly off $H = 1/3$, due to the inaccuracy of the lattice constant $a$ 
				determined in the long-wavelength condition. Error bars represent the standard 
				deviations in \textbf{a} and \textbf{b}. \textbf{c-e} Temperature dependencies 
				of the representative magnetic reflections measured at D$23$, under different 
				applied fields applied along the $c$ axis.  \textbf{c} and \textbf{d} show the 
				rocking-curve scans of the ($1/3$, $1/3$, $0.836$) reflection appearing under 
				$B$ = $0$ and the ($2/3$, $2/3$, $0$) reflection appearing under $B$ = $0.8$~T, 
				respectively, while \textbf{e} represents the reciprocal-space scan for the 
				($1/3$, $1/3$, $0.664$) peak emerging under $B = 1.3$~T along the $L$ 
				direction. r.l.u., reciprocal lattice units.; a.u., arbitrary units.
			}
			\label{EDFig8}  
		\end{figure*}
		
		% ========== Magnetic diffraction intensities ========= %
		\begin{figure*}[htp]
			\includegraphics[width=1\textwidth]{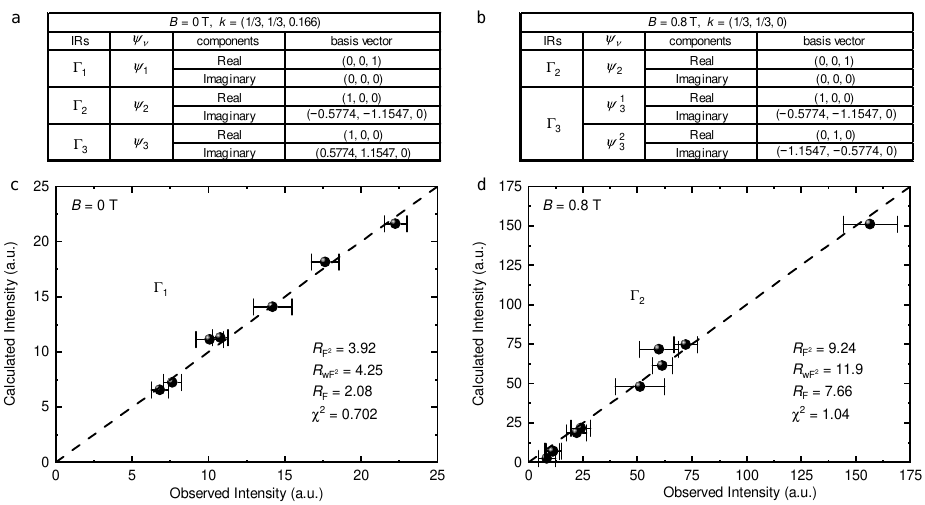}
			\caption{$\vert$ \textbf{Preliminary refinements to the magnetic diffraction intensities.}
				\textbf{a} and \textbf{b} tabulate the basis vectors of the irreducible representations (IRs) for the Co$^{2+}$ ions on the $1$b Wyckoff sites in NBCP, for \emph{\textbf{k}} $ =$ ($1/3$, $1/3$, $0.166$) and \emph{\textbf{k}} $ =$ ($1/3$, $1/3$, $0$), respectively, obtained from representation analysis. \textbf{c} and \textbf{d} show the comparisons between the observed and calculated integrated intensities of the non-equivalent magnetic reflections for $B = 0$  and $B = 0.8 $~T, adopting the modulated and non-modulated UUD spin configurations with the irreducible representation $\Gamma_{1}$ and $\Gamma_{2}$, respectively. Error bars represent the standard deviations, and the $R$ factors of the refinements are listed in both cases. a.u., arbitrary units.
			}
			\label{EDFig9}  
		\end{figure*}
		
	\end{document}